\documentclass{elsart}

\begin{document}

%remove for camera-ready copy
\begin{flushright}UMN-D-00-5 \\ July 2000 \end{flushright}

\begin{frontmatter}

%\hyphenation{ ... } 

\title{Calculations with DLCQ$^1$}%

\author{J.R. Hiller}%
%
%remove footnote for camera-ready copy
\footnotetext[1]{To appear in the proceedings of 
the Tenth International Light-Cone Meeting on
Nonperturbative QCD and Hadron Phenomenology, 
Heidelberg, June 12-17, 2000.}%
\footnote[2]{\baselineskip=14pt
Work supported in part by the Department of Energy,
contract DE-FG02-98ER41087.}

\address{Department of Physics \\
University of Minnesota Duluth \\
Duluth Minnesota 55812}

\date{24 July 2000}

\begin{abstract}
The method of discrete light-cone quantization (DLCQ) and useful
refinements are summarized.  Applications to various
field theories are reviewed. 
\end{abstract}

\end{frontmatter}

\section{Introduction}

Nonperturbative solutions are of critical importance for
the full understanding and application of quantum field
theories.  One method to obtain such solutions  is discrete
light-cone quantization (DLCQ) \cite{PauliBrodsky,review}.
A brief summary of this method, its applications, and some
refinements are given below.  The summary is much too brief
to be tutorial; an expanded discussion can be found in
Ref.~\cite{review}.  Also due to brevity, other promising
light-cone methods, such as the transverse lattice \cite{TransLattice}
and similarity transformations \cite{SimTrans}, are not discussed.

DLCQ builds from the imposition of periodic or antiperiodic boundary
conditions on a light-cone box $-L<x^-<L$, $-L_\perp<x,y<L_\perp$.
Usually couplings dictate that periodic boundary conditions
be used for bosons.  Either form can be chosen for fermions,
with antiperiodic preferred in order to avoid zero modes.  In
momentum space there is then a discrete grid:
$p^+\rightarrow n\pi/L$,
${\bf p}_\perp\rightarrow(n_x\pi/L_\perp,n_y\pi/L_\perp)$,
with $n$ even for periodic boundary conditions and odd for antiperiodic.
The limit $L\rightarrow\infty$ is exchanged for a limit
in terms of the integer {\em harmonic resolution}
$K\equiv\frac{L}{\pi}P^+$  \cite{PauliBrodsky}.
It sets the resolution for the longitudinal momentum fractions
$x=p^+/P^+\rightarrow n/K$.  The light-cone Hamiltonian
$H_{\rm LC}\equiv P^+P^-$ is independent of $L$.  Because $p^+$
is positive, the harmonic resolution limits the range of $n$ to be 
no more than $K$ and the number of constituents to be no more
than roughly $K/2$.  There is no corresponding limit for the 
transverse components, and an additional constraint must be 
imposed.  One frequently used is the invariant mass cutoff 
$m_i^2+p_{\perp i}^2\leq x_i\Lambda^2$.  The transverse integers 
then lie within some finite range, $-N_\perp\leq n_x,n_y\leq N_\perp$.

The discretization, combined with Fock-state expansions for the
eigenstates of $H_{\rm LC}$, produces a finite matrix approximation
to the infinite set of coupled integral equations for the Fock-state
wave functions.  The integrals are replaced by trapezoidal approximations
such as
\begin{equation} \int_0^1 dx \int d^2p_\perp f(x,{\bf p}_\perp)\simeq
   \frac{2}{K}\left(\frac{\pi}{L_\perp}\right)^2
   \sum_n\sum_{n_x,n_y=-N_\perp}^{N_\perp}
   f(n/K,{\bf n}_\perp\pi/L_\perp)\,,
\end{equation}
with nominal errors of order $1/K^2$ for nonsingular kernels.
Additional truncation in particle number is frequently applied, to keep
the matrix representation small.  

For large matrices the diagonalization problem is best attacked with
the Lanczos algorithm \cite{Lanczos} which requires of the matrix only
matrix-vector multiplications.  This takes good advantage of the
usually sparse structure of the matrix by allowing optimal storage
or even computation of matrix elements as needed.  An alternative
to explicit diagonalization is reduction to an effective Hamiltonian
in a single Fock sector \cite{EffH}.

\section{Brief review of calculations}

There have been a number of applications of this method.
The earliest is a nonrelativistic test \cite{NonRel},
but all the others are applications to field theories
in two or more dimensions.  The first field-theoretic
application was by Pauli and Brodsky to two-dimensional
Yukawa theory \cite{PauliBrodsky}.  Other two-dimensional
applications include 
$\phi^n$ \cite{phin}, 
QED \cite{2dQED}
(including coherent states \cite{2dCoherentStates}
and QED at finite temperature \cite{Elser}),
QCD \cite{2dQCD}, 
adjoint matter (including tube models and collinear 
models) \cite{AdjointMatter},
and the following models: 
Wick--Cutkosky \cite{2dWC}, 
sine--Gordon \cite{sineGordon}, 
Gross--Neveu \cite{GrossNeveu}, 
and Abelian Higgs \cite{AbelianHiggs}.
The behavior of the Pauli-Jordan function
$\Delta(x)=-i\int\frac{d^2k}{2\pi}
     \delta(k^2-m^2)\epsilon(k^0)e^{-ik\cdot x}$,
which should be zero for $x^2<0$, has been checked
as a test of microcausality \cite{microcausality}.
Four-dimensional applications include 
QED \cite{QED}
(including coherent states \cite{CoherentStates}),
positronium \cite{Positronium}, 
a perturbative calculation of the electron's anomalous 
moment \cite{perturbative}, 
QCD \cite{QCD},
the Wick--Cutkosky model \cite{WC}, 
a dressed fermion model \cite{PV}
(including a calculation of the $F_1$ form factor \cite{FormFactor}),
and Yukawa theory in a single-fermion truncation \cite{Yukawa}.  
The dressed fermion calculation \cite{PV} is the largest to date, having 
used basis sizes on the order of 10 million.

An important, recurring issue is that of zero modes \cite{review}.  
These take several forms: ghost fields required for a consistent
quantization, constrained fields associated with the 
periodic boundary conditions, and dynamical zero modes of
periodic gauge fields.  The zero modes are intimately tied
with the representation of what in equal-time quantization
is the structure of the vacuum.  Even if this structure is
trivial, zero modes represent corrections of order $1/K$ to 
DLCQ calculations that ignore them.  These $1/K$ effects can 
be introduced through effective interactions \cite{WC,Maeno}.

To quickly see the potential importance of zero modes, 
consider the Wick--Cutkosky model in two dimensions,
for which the interaction Lagrangian is $-g\phi|\chi|^2$.  
We impose periodic boundary conditions for $\phi$ and extract 
the zero mode field $\phi_0=\int_{-L}^L \phi dx^-/2L$.  Antiperiodic 
boundary conditions are used for the complex scalar $\chi$.  By 
integrating the equation of motion, we obtain the constraint 
$\mu^2\phi_0=-g\frac{1}{2L}\int_{-L}^L dx^-|\chi|^2$.
This implies that the interaction term of the Hamiltonian 
will include a term of the form 
$-\frac{g^2}{\mu^2}\frac{1}{2L}\left(\int_{-L}^L dx^-|\chi|^2\right)^2$,
which is unbounded from below, for any $g$, in the limit of infinite
resolution.  DLCQ calculations without zero modes can see this
unbounded spectrum \cite{2dWC}, but with some difficulty.

Recent work on zero modes includes 
chiral symmetry breaking \cite{chiSB},
symmetry breaking in $\phi^4$ \cite{SSBphi4} 
(including a treatment that does not require zero modes \cite{Thorn}),
an additional effective interaction in the massive Schwinger
model \cite{McCartor}, 
and the chiral Yukawa model \cite{chiralYukawa}.  
For earlier work, see the review in Ref.~\cite{review}.

\section{Refinements of DLCQ}

Since the original formulation of DLCQ \cite{PauliBrodsky}
several refinements and variants of the method have appeared.
Those described briefly here are 
corrections for end-point behavior \cite{vandeSande},
the use of unequal integration weights \cite{PV}, 
an indefinite-metric Lanczos algorithm \cite{Yukawa},
supersymmetric DLCQ \cite{SDLCQ,SDLCQorg}, 
and treatment of scattering amplitudes \cite{QCD00}.

When a constituent mass is small, there can be significant
end-point corrections to DLCQ wave functions.  The introduction
of an effective interaction to the DLCQ Hamiltonian can 
compensate for this \cite{vandeSande}.  The effective
interaction is constructed to include the leading
end-point behavior exactly.

Alternative integration schemes introduce unequal weights at grid
points near the boundaries \cite{Luchini,PV} to compensate for
the DLCQ grid being incommensurate with the invariant-mass cutoff.
The weights can be obtained by iterating one-dimensional integration 
rules and by picking the one-dimensional rules to satisfy chosen 
constraints, such as exact integration of linear forms.
Additional improvement can be obtained by taking into account
the cylindrical symmetry of both the integration domain and the
invariant mass constraint for two-body wave functions \cite{PV}.
The transverse integral is written in polar coordinates,
and the radial integral is approximated by a discrete sum over
the circles that intersect the points of the square grid. 

The recent calculations of the dressed fermion model \cite{PV}
and Yukawa theory \cite{Yukawa} introduce negatively normed 
Pauli--Villars particles as regulators \cite{PauliVillars}.  The 
DLCQ matrix representation is then no longer Hermitian.  Although 
there exists a Lanczos diagonalization technique for general 
matrices, an efficient special form has been developed for this 
indefinite-metric situation \cite{Yukawa}.  It produces a tridiagonal 
representation $T$ which is real and self-adjoint with respect to an 
induced indefinite metric in the Lanczos basis $\{\vec{q}_k\}$.  One 
can solve $T\vec{c}_i=\lambda_i\vec{c}_i$ for eigenvalues and right 
eigenvectors and have 
$H_{\rm LC}\vec{\phi}_i\simeq\lambda_i\vec{\phi_i}$,
with $\vec{\phi}_i=\sum_k(c_i)_k\vec{q}_k$. 

Efficient application of DLCQ to supersymmetric theories requires a 
variant, known as supersymmetric DLCQ (SDLCQ) \cite{SDLCQ}.  It is 
based on the observation \cite{SDLCQorg} that discretization of the 
supercharge $Q^-$ and computation of the Hamiltonian $P^-$ from 
the superalgebra relation $P^-=\frac{1}{2\sqrt{2}}\left\{Q^-,Q^-\right\}$
yield a discrete Hamiltonian which is explicitly supersymmetric.  The
ordinary DLCQ Hamiltonian is not supersymmetric.  The two agree only in
the limit of infinite resolution.  It is expected that zero modes 
decouple in these theories \cite{SDLCQzeromodes}, another advantage of 
supersymmetry.

An extension of DLCQ to include the calculation of scattering amplitudes
has been constructed \cite{QCD00,scattering}.  This generalizes earlier work
on the special case of $e^+e^-$ annihilation into hadrons \cite{Re+e-}.
The invariant amplitude ${\mathcal M}_{fi}$ is obtained from the
light-cone $T$ matrix, which is built from individual composite-particle
eigenstates and related operators, extending a formulation by 
Wick \cite{Wick}.

\section{Summary}

The future of DLCQ holds many exciting prospects, some of
which can already be listed.  The use of Pauli--Villars
regularization can be extended to full Yukawa theory,
QED, and perhaps QCD in the form given by Paston {\em et al.}
\cite{Paston}.  Cross sections can be calculated.  Symmetry
breaking and vacuum structure can be better understood.
For those working in string theory, higher resolution 
SDLCQ calculations in relevant theories will be of
considerable interest.

\section*{Acknowledgments}
The preparation of this review was supported in part by the 
Department of Energy, contract DE-FG02-98ER41087.

\end{document}